\documentclass[10pt,letter]{article}
\usepackage[ansinew]{inputenc}  
\usepackage{graphicx}
\usepackage{amssymb} 
\usepackage[footnotesize]{caption}
\usepackage[numbers,sort&compress]{natbib}

\newcommand{\Einheit}[1]{\ensuremath{\,\mathrm{#1}}}
\newcommand{\FracEinheit}[2]{\ensuremath{\,\frac{\mathrm{#1}}{\mathrm{#2}}}}

\newcommand{\UFReffont}{\fontsize{8pt}{10pt}\selectfont}

\newcommand{\Zw}[1]{}
	
 \newcommand{\bibformattricks}{\setlength{\itemsep}{1pt}\vspace{-30pt}}
\begin{document}
\begin{center}
\LARGE\bf All-optical Coherent Control of Electrical Currents in Single GaAs Nanowires                                        

\vspace{1.0ex}

\normalsize\bf C. Ruppert$^1$, S. Thunich$^1$, G. Abstreiter$^2$, A. Fontcuberta i Morral$^{2,3}$, A. W. Holleitner$^2$, \\ and M. Betz$^{1,4}$         

\vspace{0.5ex}

\footnotesize                																																																	
\it
$^1$ Physik-Department E11, Technische Universität München, D-85748 Garching, Germany \\
$^2$ Physik-Department and Walter Schottky Institut, Technische Universität München, D-85748 Garching, Germany \\
$^3$ Laboratoire des Matériaux Semiconducteurs, Institut des Matériaux, EPFL, CH-1015 Lausanne, Switzerland \\
$^4$ Experimentelle Physik 2, Technische Universität Dortmund, D-44221 Dortmund, Germany \\
Corresponding Author: M. Betz, e-mail address: markus.betz@tu-dortmund.de

\normalsize\rm
\vspace{0.2ex}

\end{center} 
%
%
{\setlength\leftskip{0.5in} \setlength\rightskip{0.5in}
\noindent \textbf{Abstract:}																																															 
A phase-stable superposition of femtosecond pulses and their second harmonic induces ultrashort microampere current bursts in single unbiased 
GaAs nanowires. Current injection relies on quantum interference of one- and two-photon absorption pathways.\\
\small \copyright2010 Optical Society of America\\
%
\par
}
\vspace{1.0ex}
%
%
%
%
%
\noindent 
Over the last decade, all-optical concepts to induce charge transport have sparked the interest of researchers \cite{Hach'e1997, Costa2007}. In bulk semiconductors electrical currents can be generated all-optically via a quantum interference of absorption pathways with different parity selection rules \cite{Atanasov1996}. In a particular embodiment, a phase-stable superposition of a femtosecond fundamental laser pulse ($\omega$) and its second harmonic ($2\omega$) directly injects an ultrafast current burst with a direction dictated by the pulses' polarization and relative phase. For device applications current control may be required on the single nanowire level which has not been achieved to date.

In this contribution, we experimentally demonstrate quantum interference control of ballistic electrical currents in single GaAs nanowires. Peak currents of several $\rm \mu A$ are injected with phase-related $1550 \Einheit{nm}\, (\omega)/$ $775 \Einheit{nm}\, (2\omega)$ femtosecond pulse pairs. The power and polarization dependence shows current injection to rely on a third order optical nonlinearity consistent with the established picture of interfering one- and two-photon absorption processes across the direct bandgap of GaAs.

The nanodevices of this study have been described in detail before \cite{Thunich2009}. A particular GaAs nanowire of $\sim 10 \Einheit{\mu m}$ length and $\sim 140 \Einheit{nm}$ diameter is placed across two gold pads situated $6 \Einheit{\mu m}$ apart on a $150 \Einheit{nm}$ thick SiO$_2$ layer on top of a silicon wafer. The nanowire is attached to the electrodes by a focused ion beam induced deposition of platinum. Schottky contacts between the nanowire and the gold electrodes are corroborated by the nonlinear current-voltage characteristics of the device. An scanning electron microscope (SEM) image of the device is shown in Fig. 1(a).

The optical source for current injection is a femtosecond Er:fiber laser (Toptica FFS). It delivers a $90 \Einheit{MHz}$ pulse train with pulses of $120 \Einheit{fs}$ duration and a central wavelength of $1.55 \Einheit{\mu m}$. Such telecom radiation is well suited to induce two-photon absorption 
\begin{figure}[h]
  \begin{center}
    \includegraphics[scale=1]{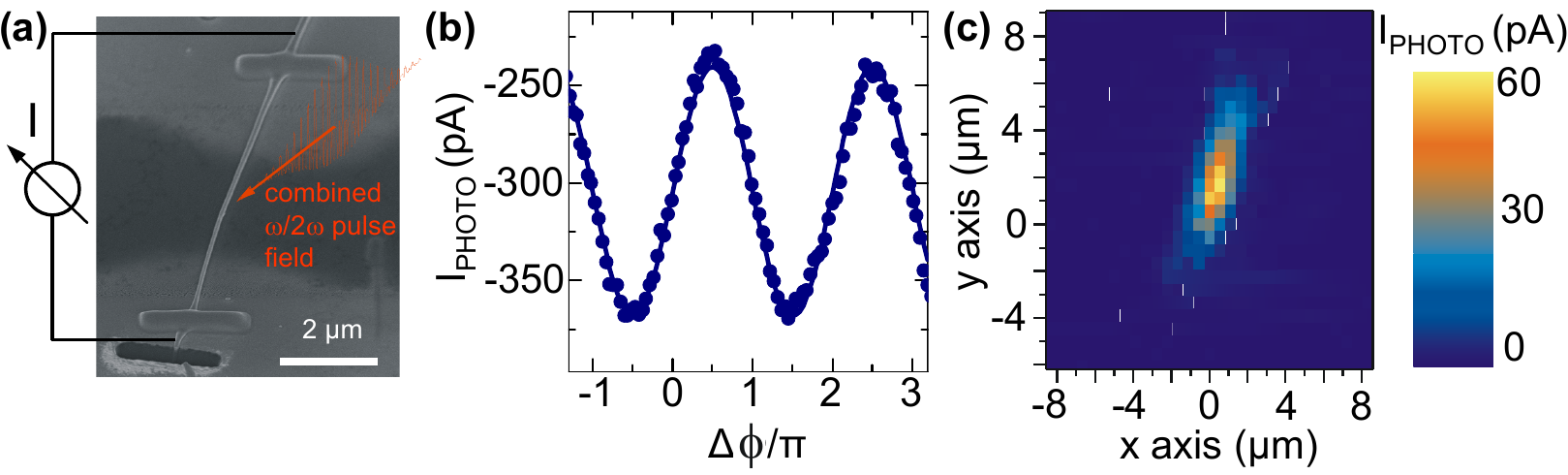}
     \setlength{\captionmargin}{0.0cm}
     \vspace{-0.5cm}
     \caption*{Fig. 1: (a) SEM image of the sample and schematic of the $\omega/2\omega$ excitation light and the photocurrent detection circuit. (b) Photoresponse $I_{PHOTO}$ of the device for various relative phases $\Delta\phi=2\phi_{\omega}- \phi_{2\omega}$ of the $\omega/2\omega$ pulse pair. From the data a phase-independent offset and the amplitude of the coherently controlled current can be extracted (cf. solid line which represents a fit to the data). (c) Two-dimensional map of the amplitude of the phase-dependent photocurrent (same scale as panel (a)).
}
     \label{figure}
  \end{center}
\end{figure}
in GaAs at room temperature ($2\cdot \hbar\omega = 1.6 \Einheit{eV} > E_G = 1.42 \Einheit{eV}$). The light is passed through a $1 \Einheit{mm}$ BiBO nonlinear optical crystal to generate several mW of second harmonic radiation at $775 \Einheit{nm}$ central wavelength. A phase-stable superposition of fundamental $(\omega)$ and second harmonic radiation $(2\omega)$ is synthesized in a two-color Michelson interferometer utilizing a dichroic beamsplitter. The photoresponse of the device is extracted with a lock-in amplifier referenced to the excitation light that is chopped at $800 \Einheit{Hz}$.
Characterizing the photoresponse of the nanodevices we find photocurrents of several $\Einheit{nA}$ at $1 \Einheit{V}$ bias voltage for peak excitation intensities of $0.02 \FracEinheit{GW}{cm^{2}}$ and $13 \FracEinheit{GW}{cm^{2}}$ of $2\omega$ and $\omega$ radiation respectively. A polarization ratio $\rho = \frac{I_{\parallel}-I_{\perp}}{I_{\parallel}+I_{\perp}}$ of $\sim 55 \, \%$ is found for both $\omega$ and $2\omega$ radiation where $I_{\parallel}$ and $I_{\perp}$ denote the photocurrent for laser polarization parallel and perpendicular to the nanowire axis. This value is comparable to previous reports and likely results from an effective attenuation of the electric field component perpendicular to the wire axis \cite{Thunich2009, Ruda2005}. 

We now turn towards the cantral aspect of this study, i.e. the quantum interference of the one- and two-photon absorbtion pathways. In bulk semiconductors it is well
known that the injection rate of a coherently controlled current is given by \cite{Hach'e1997}:	
\vspace{-0pt}
\begin{equation}\label{injection}
		\dot{J}_i = \sum_{jkl}\eta_{ijkl}E_j^{\omega}E_k^{\omega}E_l^{2\omega}\sin (2\phi_{\omega}- \phi_{2\omega}) - \frac{J_i}{\tau}
	\vspace{-0pt}
\end{equation}
\noindent where $E_{j,k,l}^{\omega,\, 2\omega}$ are the electric field amplitudes and $\phi_{\omega,\, 2\omega}$ the phases of the electromagnetic waves with frequency $\omega$ and $2\omega$. $\tau$ represents the momentum relaxation time. $\eta_{ijkl}$ is a fourth rank tensor related to the imaginary part of $\chi^{(3)}(0,\omega,\omega,-2\omega)$ \cite{Atanasov1996}. Fig. 1(b) displays the photoresponse of an unbiased nanowire for numerous relative phases $\Delta\phi=2\phi_{\omega}- \phi_{2\omega}$ within the $\omega/2\omega$ synthesized light. A phase-independent offset and an amplitude of the phase-dependent photocurrent can be extracted from the data. The latter is $62 \Einheit{pA}$ in Fig. 1(b) which, however, represents an average current flow in the external circuit. This coherently controlled electrical current is expected to rise within the duration of the $\sim 100 \Einheit{fs}$ excitation pulse and to decay with the momentum relaxation time of $\sim 200 \Einheit{fs}$. The data are obtained with a repetition rate of $90 \Einheit{MHz}$ and are, therefore, indicative of a peak current flow of $\frac{62 \Einheit{pA}}{90 \Einheit{MHz} \cdot 200 \Einheit{fs}} = 3.4 \Einheit{\mu A}$.
Spatially resolved photocurrent plots are obtained using piezo actuators to move the sample perpendicular to the optical axis. The phase-independent offset of the photoresponse has maximum amplitude when exciting the ends of the wire and is thus assumed to result from band distortions at the Schottky contacts \cite{Thunich2009}. Fig. 1(c) shows the phase-dependent signal contribution, i.e. the signature of coherently controlled currents, which has maximum signal strength when exciting the middle section of the wire. As seen from the false-color plot, however, all-optical current injection is possible at any position between the contacted ends of the wire. 

To furthermore confirm the phase-dependent photoresponse as arising from coherently controlled currents, the wire's photoresponse is analyzed as a function of the excitation strength (data not shown). The intensity dependence of the phase-dependent part upon variation of the $\omega$ ($2\omega$) is well represented by a power-law behavior with an exponent very close to 1.0 (0.5), as expected from equation (1) which predicts $\dot{J} \propto E_{2\omega}\cdot {E_{\omega}}^2$. 

Current injection also depends on the polarization of the co-polarized $\omega$ and $2\omega$ pulses with respect to the nanowire axis. The amplitude of the coherently controlled current for numerous polarization states is again extracted from phase scans similar to the one depicted in Fig. 1(b). The current has a maximum amplitude for light polarization along the wire. For light polarization perpendicular to the wire no phase-dependent current is expected and observed because in this configuration any coherently controlled current has no contribution along the wire axis. The generation of somewhat smaller currents is also possible utilizing cross-polarized $\omega$ and $2\omega$ light.

In conclusion, we have realized all-optical coherent control of femtosecond $\sim \Einheit{\mu A}$ current bursts in single semiconductor nanowires. These techniques represent the fastest way to generate charge currents and are not hampered by inductance or capacitance effects of electrical contacts. Future studies will have to focus on the optimization of charge extraction and an effective enhancement of the nonlinearity, e.g., by antenna structures. 
%
%
%

\end{document}